\documentclass[12pt]{article}
\usepackage{axodraw}

\catcode`@=12
\begin{document}
\thispagestyle{empty}
\begin{flushright} 
UCRHEP-T391\\ 
June 2005\
\end{flushright}
\vspace{0.5in}
\begin{center}
{\LARGE	\bf Hiding the Existence of a Family\\ Symmetry in the Standard 
Model\\}
\vspace{1.5in}
{\bf Ernest Ma\\}
\vspace{0.2in}
{\sl Physics Department, University of California, Riverside, 
California 92521, USA\\}
\vspace{1.5in}
\end{center}

\begin{abstract}\
If a family symmetry exists for the quarks and leptons, the Higgs sector 
is expected to be enlarged to be able to support the transformation 
properties of this symmetry. There are however three possible generic 
ways (at tree level) of hiding this symmetry in the context of the Standard 
Model with just one Higgs doublet.  All three mechanisms have their 
natural realizations in the unification symmetry $E_6$ and one in $SO(10)$. 
An interesting example based on $SO(10) \times A_4$ for the neutrino mass 
matrix is discussed.
\end{abstract}

\newpage
\baselineskip 24pt

There are three families of quarks and leptons.  The pattern of their masses 
and mixing angles has been under study for a long time.  If a family symmetry 
exists at the Lagrangian level, broken presumably only spontaneously and 
by explicit soft terms, the Yukawa couplings $f_{ijk} q_i q^c_j \phi_k$ 
and $f'_{ijk} l_i l^c_j \phi_k$ should have two or more Higgs doublets 
$\phi_k$. Otherwise $f_{ijk}$ and $f'_{ijk}$ would reduce to $f_{ij}$ 
and $f'_{ij}$.  Since the number of bilinear invariants of any given symmetry 
is very much limited, this would not result in a realistic description of 
quark and lepton mass matrices.  On the other hand, the well-tested Standard 
Model (SM) requires only one Higgs doublet (although it remains to be 
discovered experimentally).   One Higgs doublet is also preferred 
phenomenologically as an explanation of the natural suppression of 
flavor-changing neutral currents \cite{gw77}. Thus an important theoretical 
question is whether a family symmetry can be hidden in the context of the SM 
and how.  The answer is yes and there are three generic mechanisms (at tree 
level) for achieving it, as shown below. Specific new particles with masses 
well above the electroweak scale are required, but some of these exist already 
in well-known unification symmetries such as $E_6$ and $SO(10)$.

The idea is very simple.  The information concerning the family symmetry is 
encoded in the Yukawa couplings $f_{ijk}$ of quarks through the various 
$\phi_k$ Higgs doublets.  If only one Higgs doublet is allowed, the same 
information can be encoded using the dimension-five operator \cite{fn79}
\begin{equation}
{\cal L}_Y = {f_{ijk} \over \Lambda} q_i q^c_j \phi \sigma_k + H.c.,
\end{equation}
where $\sigma_k$ are heavy scalar singlets.  This mechanism is widely 
used in model building but without any discussion of how it 
may arise from fundamental interactions.  Of course, if the new interactions 
occur near the Planck scale, then they may be very strong and the effective 
operator of Eq.~(1) is nonperturbative in general.  However, if the 
new physics responsible for the family symmetry is at or below the 
quark-lepton unification scale of about $10^{16}$ GeV, then it is 
reasonable to ask how it may be realized at tree level.  In the following 
it is shown that there are three generic ways of doing this, and each will 
be discussed also in the context of the complete underlying new physics 
involved.  The assumption of this operator means that one or more of 
these generic mechanisms is likely to be correct and is thus an important 
clue to physics beyond the SM.

Even a casual observation of Eq.~(1) shows that the four fields involved 
can be grouped into the product of two pairs in only three ways, in exact 
analogy to the classic analysis of the scattering of two particles into 
two particles.  The intermediate states must then have well-defined 
transformation properties under the standard $SU(3)_C \times SU(2)_L \times 
U(1)_Y$ gauge group.  These are depicted in Figures 1 to 3.  The heavy 
quarks $Q_{1,2}$ and $Q^c_{1,2}$ are $SU(2)_L$ singlets (doublets) and $H$ 
are heavy scalar doublets.

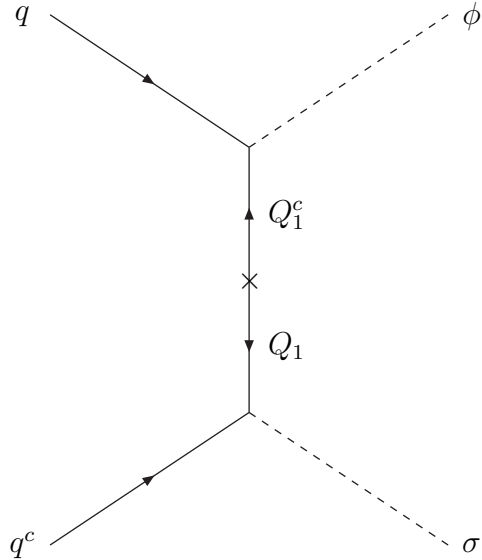
\begin{figure}[htb]
\begin{center}
\begin{picture}(400,225)(0,0)
\ArrowLine(125,200)(200,150)
\ArrowLine(200,100)(200,150)
\ArrowLine(200,100)(200,50)
\ArrowLine(125,0)(200,50)
\DashLine(275,200)(200,150)3
\DashLine(275,0)(200,50)3
\Text(115,200)[]{$q$}
\Text(115,0)[]{$q^c$}
\Text(215,125)[]{$Q^c_1$}
\Text(201,100)[]{$\times$}
\Text(215,75)[]{$Q_1$}
\Text(285,200)[]{$\phi$}
\Text(285,0)[]{$\sigma$}
\end{picture}
\end{center}
\caption{Realization of Eq.~(1) with heavy $Q_1$ and $Q^c_1$ singlets.}
\end{figure}

Consider first Fig.~1.  Under $SU(3)_C \times SU(2)_L 
\times U(1)_Y$, $q \sim (3,2,1/6)$ and $\phi \sim (1,2,\pm1/2)$, hence 
$Q^c_1 \sim (3^*,1,-2/3)$ or $(3^*,1,1/3)$ is required.  This means 
$Q_1 \sim (3,1,2/3)$ or $(3,1,-1/3)$, i.e. heavy quark singlets with 
charges equal to either those of the $u$ quarks or $d$ quarks.  This 
mechanism is thus equivalent to that of the canonical seesaw mechanism for 
Dirac fermions \cite{dwr87}.  The effective family structure of Eq.~(1) is 
then given by
\begin{equation}
{f_{ijk} \over \Lambda} = y_{ia} (M^{-1})_{ab} h_{bjk},
\end{equation}
where $y_{ia}$ are the couplings of $q_i (Q^c_1)_a \phi$, $h_{bjk}$ those 
of $(Q_1)_b q^c_j \sigma_k$, and $M$ the mass matrix of $Q_1 Q^c_1$.  
Since the family symmetry applies to all three of these quantities, it is 
well hidden in the resulting effective operator of Eq.~(1) and even more 
so in the resulting mass matrix
\begin{equation}
m_{ij} = {f_{ijk} \over \Lambda} \langle \sigma_k \rangle \langle \phi 
\rangle.
\end{equation}
On the positive side, if this particular 
mechanism is assumed, specific models of family structure may be considered 
and then compared to the data.  Singlet quarks of charge $-1/3$ are 
contained in the fundamental \underline {27} representation of $E_6$.  
Hence the $d$ quarks of the SM may owe their family structure wholly or 
partly \cite{r00} to such a mechanism.

\begin{figure}[htb]
\begin{center}
\begin{picture}(400,225)(0,0)
\ArrowLine(125,200)(200,150)
\ArrowLine(200,100)(200,150)
\ArrowLine(200,100)(200,50)
\ArrowLine(125,0)(200,50)
\DashLine(275,200)(200,150)3
\DashLine(275,0)(200,50)3
\Text(115,200)[]{$q$}
\Text(115,0)[]{$q^c$}
\Text(215,125)[]{$Q^c_2$}
\Text(201,100)[]{$\times$}
\Text(215,75)[]{$Q_2$}
\Text(285,200)[]{$\sigma$}
\Text(285,0)[]{$\phi$}
\end{picture}
\end{center}
\caption{Realization of Eq.~(1) with heavy $Q_2$ and $Q^c_2$ doublets.}
\end{figure}
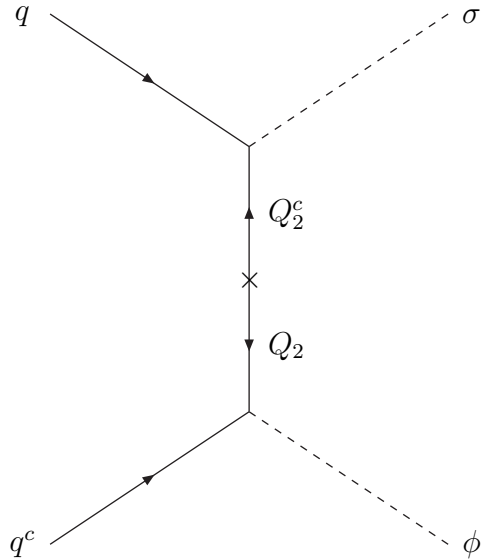

Consider next Fig.~2.  Here $Q_2 \sim (3,2,1/6)$ and $Q^c_2 \sim 
(3^*,2,-1/6)$ are required.  Whereas these heavy vector quark doublets are 
not present in the \underline {27} of $E_6$, the corresponding heavy lepton 
doublets $L_2 \sim (1,2,-1/2)$ and $L^c_2 \sim (1,2,1/2)$ are, and 
they have been used for example in a recently proposed model 
\cite{hmp05} of late neutrino mass and baryogenesis.  Thus the observed 
lepton family structure may be encoded with this mechanism. 

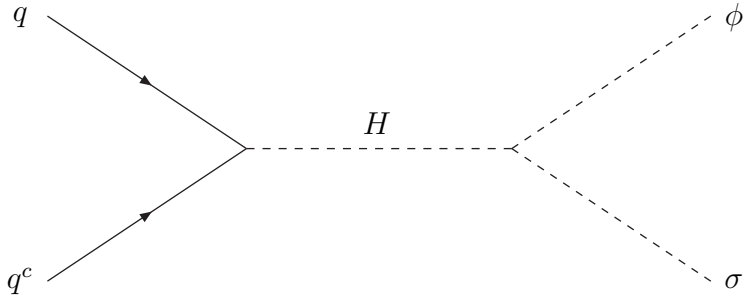
\begin{figure}[htb]
\begin{center}
\begin{picture}(400,125)(0,0)
\ArrowLine(75,100)(150,50)
\ArrowLine(75,0)(150,50)
\DashLine(325,100)(250,50)3
\DashLine(325,0)(250,50)3
\DashLine(150,50)(250,50)3
\Text(65,100)[]{$q$}
\Text(65,0)[]{$q^c$}
\Text(200,60)[]{$H$}
\Text(335,0)[]{$\sigma$}
\Text(335,100)[]{$\phi$}
\end{picture}
\end{center}
\caption{Realization of Eq.~(1) with heavy scalar $H$ doublets.}
\end{figure}

Finally consider Fig.~3.  The analog of Eq.~(2) is
\begin{equation}
{f_{ijk} \over \Lambda} = h_{ija} (M^2)^{-1}_{ab} \mu_{bk},
\end{equation}
where $h_{ija}$ are the couplings of $q_i q^c_j H_a$, $\mu_{bk}$ those of 
$H^\dagger_b \phi \sigma_k$, and $M^2$ the mass-squared matrix of $H$. This 
mechanism is realized naturally for example in $SO(10)$ (as well as $E_6$), 
where $q$ and $q^c$ are $SU(2)_L$ and $SU(2)_R$ doublets respectively, $H$ 
the heavy scalar bidoublets which carry the family structure, and $\phi$ the 
SM scalar doublet. It is appplicable to all Dirac fermions, including the $u$ 
quarks. It differs from the usual realization of quark masses in left-right 
gauge models where $H$ is a scalar bidoublet at the electroweak scale.

Since $q q^c$ couples to $\phi \sigma$ through $H$ in Fig.~3, the full 
Higgs potential involving all 3 scalar fields should be considered. 
As a simple example, consider the case where $\sigma$ and $H$ transform 
in the same way under an extra $U(1)$ symmetry but $\phi$ is trivial, 
so that $H^\dagger \phi \sigma$ is an allowed term in the Lagrangian 
but $H^\dagger \phi$ is not.  The most general Higgs potential involving 
$\sigma$, $H$, and $\phi$ is then given by
\begin{eqnarray}
V &=& m_\sigma^2 \sigma^\dagger \sigma + m_H^2 H^\dagger H + m_\phi^2 
\phi^\dagger \phi + {1 \over 2} \lambda_1 (\sigma^\dagger \sigma)^2 + 
{1 \over 2} \lambda_2 (H^\dagger H)^2 + {1 \over 2} \lambda_3 
(\phi^\dagger \phi)^2 \nonumber \\ &+& \lambda_4 (\sigma^\dagger \sigma)
(H^\dagger H) + \lambda_5 (\sigma^\dagger \sigma)(\phi^\dagger \phi) 
+ \lambda_6 (H^\dagger H)(\phi^\dagger \phi) + \lambda_7 (H^\dagger \phi)
(\phi^\dagger H) \nonumber \\ &+& [\mu H^\dagger \phi \sigma + H.c.]
\end{eqnarray}
Let $\mu$ be real, as well as $\langle \sigma \rangle = x$, $\langle H 
\rangle = u$, and $\langle \phi \rangle = v$.  Then the minimization of 
$V$ results in the 3 conditions:
\begin{eqnarray}
&& x[m_\sigma^2 + \lambda_1 x^2 + \lambda_4 u^2 + \lambda_5 v^2] + 
\mu u v = 0, \\
&& u[m_H^2 + \lambda_2 u^2 + \lambda_4 x^2 + (\lambda_6 + \lambda_7)v^2] + 
\mu v x = 0, \\
&& v[m_\phi^2 + \lambda_3 v^2 + \lambda_5 x^2 + (\lambda_6 + \lambda_7)u^2] + 
\mu u x = 0.
\end{eqnarray}
Since $u,v << x$ is required for electroweak symmetry breaking,
\begin{equation}
x^2 \simeq {-m_\sigma^2 \over \lambda_1}
\end{equation}
is obtained from Eq.~(6).  Assuming now that $m_H^2 + \lambda_4 x^2 > 0$, 
Eq.~(7) then yields
\begin{equation}
u \simeq {-\mu v x \over m_H^2 + \lambda_4 x^2}.
\end{equation}
Substituting the above into Eq.~(8), the following effective condition 
for $v$ is obtained:
\begin{equation}
m_\phi^2 + \lambda_5 x^2 - {\mu^2 x^2 \over m_H^2 + \lambda_4 x^2} + 
\left[ \lambda_3 + {(\lambda_6 + \lambda_7) \mu^2 x^2 \over (m_H^2 + \lambda_4 
x^2)^2} \right] v^2 = 0.
\end{equation}
Using Eqs.~(6) to (8), the mass-squared matrix spanning the neutral real 
components of $\sigma$, $H$, and $\phi$ is given by
\begin{equation}
{\cal M}^2_{\sigma,H,\phi} = \pmatrix{2 \lambda_1 x^2 - \mu u v/x & 
2 \lambda_4 x u + \mu v & 2 \lambda_5 x v + \mu u \cr 2 \lambda_4 x u + 
\mu v & 2 \lambda_2 u^2 - \mu v x/u & 2 (\lambda_6 + \lambda_7) u v + 
\mu x \cr 2 \lambda_5 x v + \mu u & 2 (\lambda_6 + \lambda_7) u v + 
\mu x & 2 \lambda_3 v^2 - \mu u x/v}.
\end{equation}
Since $x >> u,v$, two approximate eigenstates are $\sigma$ and 
$(v H - u \phi)/\sqrt{v^2+u^2}$ with $m^2 \simeq 2 \lambda_1 x^2$ and 
$ -\mu x(v^2+u^2)/vu$ respectively.  Thus all scalar fields 
are heavy except for the linear combination 
$(v \phi + u H)/\sqrt{v^2+u^2}$ which is identical to the single Higgs 
doublet of the SM.  If the latter is extended to include supersymmetry, 
then there will be two Higgs doublets, as in the MSSM (Minimal 
Supersymmetric Standard Model).

In the mechanism of Fig.~3, it is clear that the $q q^c$ mass matrix may 
also be written as
\begin{equation}
m_{ij} = h_{ijk} \langle H_k \rangle,
\end{equation}
where $\langle H_k \rangle$ is given by the generalization of Eq.~(10). 
The family structure is determined not only by $h_{ijk}$ which may come 
from an assumed symmetry, but also by $\langle H_k \rangle$ which is 
hidden in the dynamics of the scalar sector much above the electroweak 
scale.  However, if the family symmetry is global, and broken only 
spontaneously, then a massless Goldstone boson, the familon, will 
appear \cite{w82}.

As an application of the mechanism of Fig.~3, consider the non-Abelian 
discrete symmetry $A_4$, the group of the even permutation of 4 objects 
which is also the symmetry group of the tetrahedron.  It has been 
discussed \cite{mr01,bmv03,m04,af05,m05} as a family symmetry for 
the understanding of the neutrino mass matrix.  Suppose it is 
combined with $SO(10)$.  Then all quarks and leptons are naturally assigned 
as $({\bf 16}; \underline{3})$ under $SO(10) \times A_4$.  [There 
are 3 inequivalent irreducible singlet representations of $A_4$, 
$\underline{1}$, $\underline{1}'$, $\underline{1}''$, and 1 irreducible 
triplet representation $\underline{3}$.]  This assignment differs from the 
original one \cite{mr01,bmv03} where $q,l \sim \underline{3}$ but $q^c,l^c 
\sim \underline{1}, \underline{1}', \underline{1}''$, which cannot be 
embedded into $SO(10)$.  The heavy scalar $H$ should then be assigned as 
$(\overline{\bf 10}; \underline{1}, \underline{1}', \underline{1}'')$, 
$\sigma$ as $(\overline{\bf 16}; \underline{1}, \underline{1}', 
\underline{1}'')$, and $\phi$ as $(\overline{\bf 16}; \underline{1})$.  
For $a_{1,2,3} \sim \underline{3}$ and $b_{1,2,3} \sim 
\underline{3}$ under $A_4$,
\begin{eqnarray}
&& a_1 b_1 + a_2 b_2 + a_3 b_3 \sim \underline{1}, \\
&& a_1 b_1 + \omega^2 a_2 b_2 + \omega a_3 b_3 \sim \underline{1}', \\
&& a_1 b_1 + \omega a_2 b_2 + \omega^2 a_3 b_3 \sim \underline{1}'',
\end{eqnarray}
where $\omega = e^{2 \pi i/3}$.  Hence all quark and lepton Dirac mass 
matrices are diagonal but with arbitrary eigenvalues.  This is actually 
a rather good approximation in the quark sector, where all mixing 
angles are known to be small.  In fact, if the theory is also 
supersymmetric, then the explicit breaking of $A_4$ in the soft 
supersymmetry-breaking terms themselves could be used to generate 
a realistic quark mixing matrix \cite{bdm99}.  In the lepton sector, the 
same could be accomplished \cite{hrsvv04} in the case of the BMV 
model \cite{bmv03}.  On the other hand, in the context of $SO(10)$, 
the $\overline{\bf 126}$ representation can be used to obtain Majorana 
neutrino masses according to the well-known seesaw formula \cite{seesaw}
\begin{equation}
{\cal M}_\nu = {\cal M}_L - {\cal M}_D {\cal M}_R^{-1} {\cal M}_D^T,
\end{equation}
where
\begin{equation}
{\cal M}_D = \pmatrix{x & 0 & 0 \cr 0 & y & 0 \cr 0 & 0 & z}.
\end{equation}
As for ${\cal M}_{L,R}$, using the assignment $(\overline{\bf 126};
\underline{3})$, they are both naturally of the form
\begin{equation}
{\cal M}_L = \pmatrix{0 & d & d \cr d & 0 & d \cr d & d & 0}, ~~~
{\cal M}_R = \pmatrix{0 & D & D \cr D & 0 & D \cr D & D & 0}.
\end{equation}
Thus
\begin{equation}
{\cal M}_R^{-1} = {1 \over 2D} \pmatrix{-1 & 1 & 1 \cr 1 & -1 & 1 \cr 
1 & 1 & -1},
\end{equation}
and
\begin{equation}
{\cal M_D} {\cal M}_R^{-1} {\cal M}_D^T = {1 \over 2D} \pmatrix{-x^2 & xy & 
xz \cr xy & -y^2 & yz \cr xz & yz & -z^2} = - \pmatrix{a & b & c \cr 
b & b^2/a & -bc/a \cr c & -bc/a & c^2/a},
\end{equation}
which has 3 zero $2 \times 2$ subdeterminants as expected \cite{ma05}. 
Together with ${\cal M}_L$, this becomes a four-parameter hybrid description 
\cite{cfm05} of the neutrino mass matrix.  Let $b=c$, then in the basis 
spanning $\nu_e$, $(\nu_\mu+\nu_\tau)/\sqrt{2}$, and 
$(-\nu_\mu+\nu_\tau)/\sqrt{2}$,
\begin{eqnarray}
{\cal M}_\nu = \pmatrix{a & \sqrt{2} (d+b) & 0 \cr \sqrt{2} (d+b) 
& d & 0 \cr 0 & 0 & -d+2b^2/a},
\end{eqnarray}
which is a new and very interesting pattern. It implies that $\theta_{13} = 0$ 
and $\theta_{23} = \pi/4$ in the neutrino mixing matrix, in agreement with 
data.  Assuming $d,a,b$ to be real, the solar mixing angle is given by
\begin{equation}
\tan 2 \theta_{12} = {2 \sqrt{2} (d+b) \over d-a},
\end{equation}
which yields $\tan^2 \theta_{12} = 0.5$ in the limit $a=b=0$, again in 
agreement with data and realizing the so-called tri-bimaximal mixing 
pattern suggested \cite{hps02} some time ago.  Assuming thus that
\begin{equation}
a << b << d << b^2/a,
\end{equation}
the 3 neutrino mass eigenvalues in this case are approximately $-d$, $2d$, 
and $2b^2/a$, resulting in
\begin{equation}
\Delta m^2_{sol} \simeq 3d^2, ~~~ \Delta m^2_{atm} \simeq 4b^4/a^2.
\end{equation}
As a numerical example, let
\begin{equation}
a = 2.6 \times 10^{-5}~{\rm eV}, ~~~ b = -8.5 \times 10^{-4}~{\rm eV}, 
~~~ d = 5.5 \times 10^{-3}~{\rm eV},
\end{equation}
then
\begin{equation}
\tan^2 \theta_{12} = 0.45, ~~~ \Delta m^2_{sol} = 7.9 \times 10^{-5}
~{\rm eV}^2, ~~~ \Delta m^2_{atm} = 2.4 \times 10^{-3}~{\rm eV}^2.
\end{equation}
More details of this model will be presented elsewhere.

In conclusion, it has been shown how the Standard Model with one Higgs 
doublet may hide the existence of a family symmetry for the quarks and 
leptons.  The new heavy particles involved in 3 tree-level realizations of 
this mechanism have been identified and found to be available in the 
unification symmetries $E_6$ and $SO(10)$.  A new and very interesting 
model based on $SO(10) \times A_4$ for the neutrino mass matrix is 
obtained.

This work was supported in part by the U.~S.~Department of Energy under 
Grant No. DE-FG03-94ER40837.

\newpage
\bibliographystyle{unsrt}

\end{document}